\documentclass[sn-mathphys,Numbered]{sn-jnl}% Math and Physical Sciences Reference Style
\usepackage{graphicx}%
\usepackage{multirow}%
\usepackage{amsmath,amssymb,amsfonts}%
\usepackage{amsthm}%
\usepackage{mathrsfs}%
\usepackage[title]{appendix}%
\usepackage{xcolor}%
\usepackage{textcomp}%
\usepackage{manyfoot}%
\usepackage{booktabs}%
\usepackage{algorithm}%
\usepackage{algorithmicx}%
\usepackage{algpseudocode}%
\usepackage{listings}%
\usepackage{url}
\begin{document}

\title{Topics evolution through multilayer networks}
\subtitle{Analysing 2M tweets from 2022 Qatar FIFA World Cup}


\author*[1,2]{\fnm{Andrea} \sur{Russo}}\email{andrea.russophd@gmail.com}

\author[1]{\fnm{Vincenzo} \sur{Miracula}}\email{vincenzo.miracula@phd.unict.it}
\equalcont{These authors contributed equally to this work.}

\author[1]{\fnm{Antonio} \sur{Picone}}\email{antonio.picone@phd.unict.it}
\equalcont{These authors contributed equally to this work.}

\affil*[1]{\orgdiv{Department of Physics and Astronomy "E. Majorana"}, \orgname{University of Catania}, \orgaddress{\street{via S. Sofia}, \city{Catania}, \postcode{95123}, \state{Italia}, \country{Sicilia}}}

\affil*[2]{\orgdiv{Sorbonne University}, \orgname{Lettres Sorbonne Universite}, \orgaddress{\street{1 Rue Victor Cousin}, \city{Paris}, \postcode{75005}, \state{France}, \country{Ile de France}}}


\abstract{In this study, we conducted a comprehensive data collection on the 2022 Qatar FIFA World Cup event and used a multilayer network approach to visualize the main topics, while considering their context and meaning relationships. We structured the data into layers that corresponded with the stages of the tournament and utilized Gephi software to generate the multilayer networks. Our visualizations displayed both the relationships between topics and words, showing the word-context relationship, as well as the dynamics and changes over time by layer of the most frequently discussed topics.}

\maketitle

\section{Introduction} 
\label{intro}
In complexity sciences, due to the massive amount of data collected from the sample under observation and the differences in the data collected, a simple layered network is inadequate to express and illustrate the interactions of numerous interdependent components in a whole system.
Such systems, and the self-organization and emergent phenomena they exhibit, are central to many important global challenges for the future of the worldwide knowledge society \cite{boccaletti2014structure}.
Therefore, in this paper, we have chosen to create a multilayer network \cite{PhysRevX.3.041022} to demonstrate how, in a social sample, individuals may change topics based on various timing events and interactions within a given event. We believe that multilayer networks offer a more nuanced and comprehensive representation of social contexts, allowing for a deeper understanding of social dynamics, events, and interactions.

For example, multilayer networks are valuable for understanding social event interaction because they can encompass various social scales, providing a comprehensive overview of social structure \cite{finn2021multilayer}. They reveal hidden patterns that are not visible in single-layer networks \cite{sayles2019social}, facilitate the study of complex social interactions and their social evolution \cite{sayles2019social, mourier2019multilayer}, and ultimately aid in the understanding of factors influencing social behavior \cite{battiston2020networks}. 
Our objective is to highlight the potential of multilayer networks, not just for comprehending individual social interactions, but also for unraveling how entire social events give rise to emergent behaviors that would otherwise be challenging to discern through interaction analysis alone, primarily because the event itself is often overlooked.

The FIFA World Cup in Qatar in 2022 was a great opportunity to collect a huge amount of data about various topics and fans supporting national teams.
We have collected almost two million (1,923,283) tweets from the 2022 Qatar FIFA World Cup (the entire event counted 2.2 million tweets)\footnote{https://getdaytrends.com/trend/\%23FIFAWorldCup/}, with the hashtags ``\#FIFAWorldCup'' and ``\#QATAR2022''.

In order to assess the significance of an event within various social dynamics, it is necessary to analyze and gather information from the ``context'' of the event. Examining the context helps us comprehend the importance of specific topics and related words, enabling us to determine the relative importance of certain topics over others.

In this paper, therefore, we attempted to compare the differences between various layers within a single multi-timing event. Our goal was to gather more information about the ``context'' of the most significant topics, which can reveal emerging social behaviors.

\section{Data \& Method} 
\label{method}
We collected the Twitter data using the Twitter API. To accurately represent the desired information and create the multilayer network \cite{kivela2014multilayer}, we started by analyzing the data and then divided the database according to the time of data collection. This allowed us to obtain temporally correct data for each stage of the tournament and visualize it in 3D \cite{mcgee2019state}.
The dates for the tournament stages are listed in the Table \ref{tab:1}.

\begin{table}[htb] % [htb]
    \centering\small
    \scalebox{1}{
    \begin{tabular}{l|c|c}
    \hline
Stages & Start dates (2022) & Ending dates (2022)\\
             \hline
         Group stage & 20 November   & 2 December  \\
         Round of 16  & 3 December &  6 December \\
         Quarter-finals & 9 December & 10 December \\
         Semi-finals& 13 December &  14 December\\
         Final  & 17 December & 18 December\\
         \hline
    \end{tabular}}
    \caption{Dates used by Stages-Layers}
    \label{tab:1}
\end{table}

After separating the various layers, we cleaned the dataset using stop-words. Then, to obtain a word network that could depict the topics, as well as the context and connected meaning of them over time, we analyzed the data using an algorithm called Bigram.

A bigram is a sequence of two adjacent elements from a string of tokens, which are typically letters, syllables, or words. A bigram is an n-gram for n = 2. The frequency distribution of every bigram in a string is commonly used for simple statistical analysis of text in many applications, including computational linguistics, cryptography, speech recognition, and so on. We did not use \textit{TD-IDF} or a similar algorithm because we opted for Bigram as the best software for our words-context relation goal. In our code, we selected the most used words by users (nodes) and linked them (edges) with the second most used words in the same sentence.

We selected the most frequently occurring words for each time layer. After multiple experiments\footnote{\url{https://github.com/AndreaRussoAgid/Multilayer_FRCCS_2023.git}}, we decided to set a limit of 300 words. This choice strikes a balance between Gephi's technical limitations\footnote{\url{https://answers.launchpad.net/gephi/+question/107399}} and the challenge presented by a large number of words with numerous nodes and edges. Too many nodes and edges can hinder comprehension and fail to offer useful information such as the text's context.

To express the concept of each topic, we arranged the words with greater weight, akin to the topics, linked to those with less weight. This strategic linkage allows us to extract not only the primary topic but also the associated words that collectively convey an opinion, thereby providing insights into both the event and the topic itself. Context, defined as \textit{the parts of something written or spoken that immediately precede and follow a word or passage and clarify its meaning} \cite{Looking_At_The_2023_cambridge}, represents the type of information that may naturally evolve over time. Nevertheless, it emphasizes and clarifies the reasons, motivation, and dynamics of why that event/topic is being discussed, providing the entire interaction.

Regarding the multilayer, we used ``World'', ``fifa'', and ``Team'' as the key pillar-words for all layers. There are several tools available for visually representing multilayers, including Pymnet\footnote{\url{http://www.mkivela.com/pymnet/}}, but we believe that Gephi is certainly superior at the graphical level.

It is a bit difficult with Gephi despite having MultiViz \cite{MultiViz} to differentiate layers. In fact, even though this tool provides the option to create multi-layers based on certain parameters, it is not easy to generate a multi-layer from the obtained data due to the repetition of many words at both the Nodes and Edges levels in the different layers.

To address this issue, we modified the word ID by adding symbols that indicate both the layer and the reference edges between nodes. For the Group stage, we did not include any symbols. However, for the other stages (for the purpose of recognition described above), we chose the following symbols: `` \^{} '' for the Round of 16, `` * '' for the quarters, `` \dag '' for the semifinals, and `` \ddag '' for the finals.

In conclusion, we utilized the Gephi community algorithm (modularity class) to explore ``word communities'' and consequently enhance the context and significance of the reference hub and its related words.

\section{Results \& Conclusion} 
\label{res}

After performing the cleaning and bigram process, and excluding nodes and edges that are not part of the main network (gigantic component), we obtained a multilayer network with 858 nodes and 1041 edges.

\begin{figure}
    \centering
    \includegraphics[width=12cm,height=17cm]{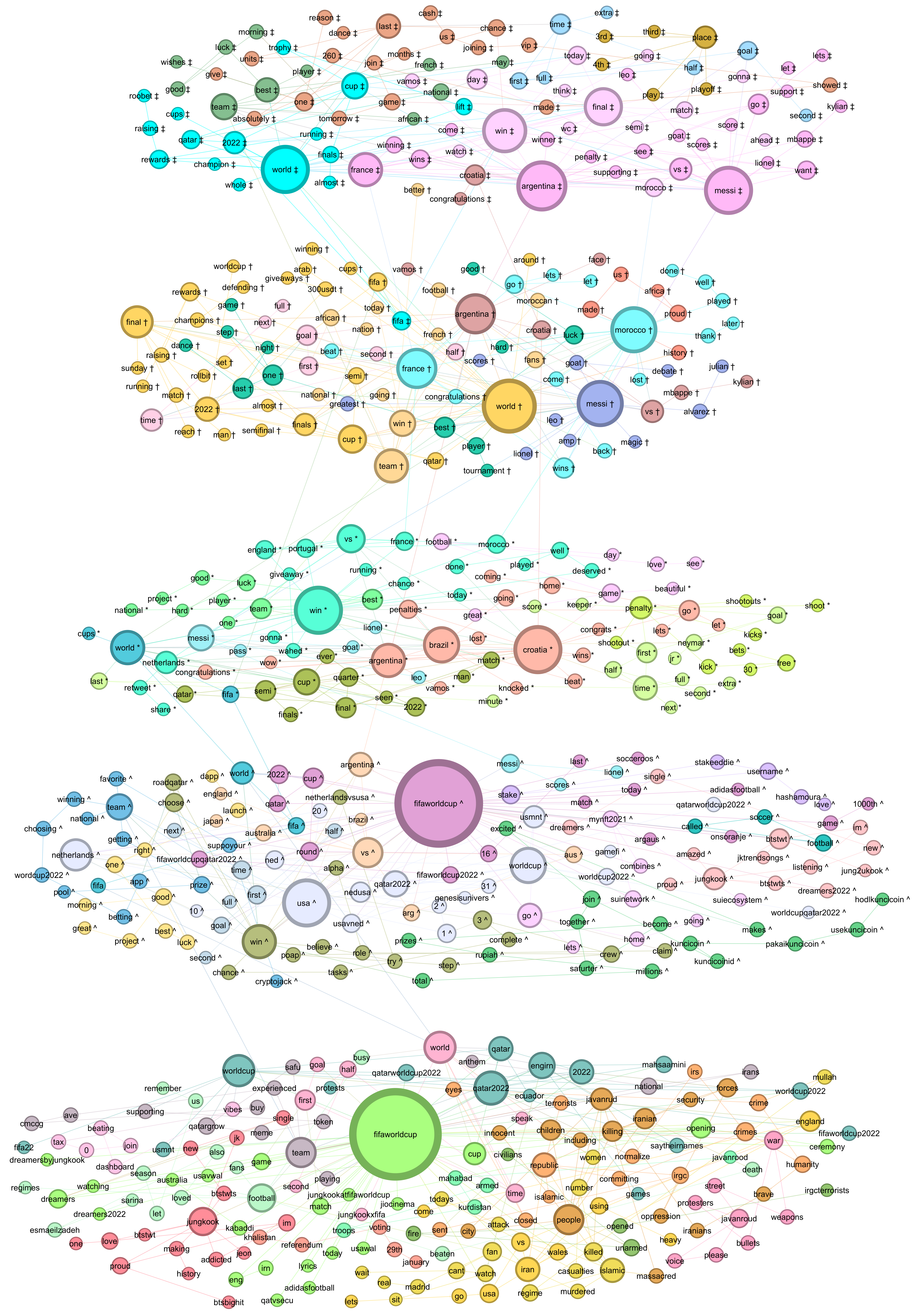} 
    \caption{Topics Multilayer networks about 2022 Qatar FIFA World Cup}
    \label{fig:1}
\end{figure}

The network shown in the Figure \ref{fig:1} illustrates the changing topics of discussion across different layers. Through this depiction, the overall structure and dynamics of the event become visible, while emergent behaviors influenced by the context are mainly observed within each individual layer. This was further supported by Gephi's community algorithm, which highlighted distinct communities using different colors.

During the Group stage, word connections related to crime and social injustices (orange and yellow) and the ceremony's performer, the \textit{BTS} group (in red), were observed creating a context for discussion. Contrary to common expectations, there are no significant communities within this layer discussing the tournament itself, football, or football teams.

However, as we progressed to the subsequent stages (Round of 16, Quarters, Semifinals, and Finals), there was a shift in the focus of the discussion, with less emphasis on the overall event and more on specific events, teams, and players.

For example, during the Round of 16 `` \^{} '', we observe a consistent presence of discussions related to BTS (salmon pink), but we also uncover conversations regarding the teams (light blue) that will advance to the Round of 16. Interestingly, some of the tournament's top seeds, including Germany and Belgium, did not progress, while anticipated matchups like USA vs. Netherlands (white smoke) are being actively discussed.

During the quarter-finals `` * '', we observed communities engaging in discussions related to refereeing (yellow line). There have been controversies surrounding the addition of excessive minutes to extra time in each match, which some believe is aimed at drawing a larger audience. These discussions have also polarized the community regarding penalties and fouls \cite{Ben_Morse_2023_cnn, Dorian_Geiger_2023_aljazeera}. 
Furthermore, within the same layer, we noticed the initiation of discussions and betting activities (aqua green) focused on predicting the winning team in the football competition.

Regarding the semifinals `` \dag '', the connected set of words notably emphasizes a context of appreciation and gratitude towards the Moroccan players (cyan and bright orange), acknowledging their remarkable achievement in reaching the semifinals and competing in the 3-4 place final. It is also evident (light purple and dark pink) that the upcoming challenge between Messi and Mbappe is beginning to be discussed (which will also be featured in the final), and how the polarization (sand yellow, yellow, and cyan) has led to a discussion about the final showdown between France and Argentina.

Lastly, in the final layer `` \ddag '', Twitter users have engaged in discussions (represented by dark pink) concerning the intense rivalry between Messi and Mbappe during the final. This has led to a clear polarization of opinions (depicted by the interplay of light pink and dark pink). Additionally, there has been a substantial volume of conversation (indicated by dark orange) revolving around Messi's \textit{last dance}, referring to the potential conclusion of his national team career, driven by both his age and past retirement announcements.

\section{Conclusion}
The scientific community is increasingly turning to the utilization of layers and multilayer networks as a means to comprehend the complexities of social order and dynamics. While multilayer networks provide a promising avenue for researchers to delve into social dynamics, the realm of understanding social events themselves and their evolution over time remains relatively uncharted.

Our work, however, aims to bridge this gap by highlighting the potential of multilayers and network science in unraveling the complexities of evolving social events. Nevertheless, this undertaking presented both challenges and opportunities. The creation of the multilayer structure posed challenges as the analysis of interactions across multiple layers over time is still in its early stages, lacking suitable tools for seamless implementation.
On the other hand, we discovered opportunities in exploring the contextual dimensions of events, which enabled us to extract invaluable insights. This contextual information is essential in comprehending the various interactions among network users and understanding the underlying motivations driving these interactions. Consequently, we attain a deeper understanding of the intricate tapestry of social dynamics and the potential emergence of new behaviors within them.

In conclusion, our work demonstrates:

\begin{enumerate}
\item The capacity to create multilayer networks that showcase the evolution of events and social dynamics over time.
\item We also emphasize the ability to acquire valuable contextual information, clarifying the reasons behind interpersonal interactions.
\item Our methodology enables the effective visualization of complete social events through multilayer network analysis techniques.
\end{enumerate}

\section{Acknowledgments}

This accomplishment was made possible through an extensive data collection effort from Twitter, which is available upon request due to its substantial size (exceeding several gigabytes). Additionally, we devoted ongoing efforts to optimize the network on multiple fronts. This optimization aimed to achieve an ideal visualization level that enables human observers to extract meaningful information from the network. Our endeavors included conducting various tests, with a limit exceeding 300 nodes, to fine-tune this aspect. Furthermore, we tackled the challenge of creating the network itself, recognizing that Gephi, the tool we utilized, was not originally designed with multilayers in mind.
The various network tests can be observed at the following link: \url{https://github.com/AndreaRussoAgid/Multilayer_FRCCS_2023.git}.

\section{Funding}
The author(s) disclosed receiving the following financial support for the research, authorship, and/or publication of this article:
This project has received funding from the University of Catania.

\section{Author Contributions}
Investigation, Data elaboration, and Methodology: A.R.; Data resources: V.M.; Data cleaning and Software: A.P. and V.M.; Correction: A.P. All authors have read and agreed to the published version of the manuscript.

\bibliography{biblio}  

\end{document}